\documentclass[12pt]{article}
\pdfoutput=1
\usepackage{latexsym}

\usepackage{graphicx}

\def\tr{{\rm tr}}
\def\ket#1{\mid~\!\!\!{#1}~\!\!\rangle}

\def\qm{quantum mechanics}
\def\QMl{quantum mechanical }

\def\M{measurement }
\def\m{measurement}
\def\${\enskip$}

\begin{document}

{\bf \large \noindent  On EPR-type
Entanglement in the Experiments of
Scully et Al. II. Insight in the Real
Random\\ Delayed-choice
Erasure Experiment}\\

{\bf \noindent F. Herbut}\\

\vspace{0.5cm}

\rm

\noindent {\bf Abstract} It was
pointed out in the first part of
this study [Herbut:Found. Phys.
{\bf 38}, 1046-1064 (2008)]
 that
EPR-type entanglement is defined
by the possibility of performing
any of two mutually incompatible
distant, i. e.,
direct-interaction-free, \m s.
They go together under the term
'EPR-type disentanglement'. In
this second part,
quantum-mechanical insight is
gained in the real random
delayed-choice erasure experiment
of Kim et al. [Kim et al.: Phys.
Rev. Lett. {\bf 84}, 1-5 (2000)]
by a
relative-reality-of-unitarily-evolving-state
(RRUES) approach (explained in the
first part). Finally, it is shown
that this remarkable experiment,
which performs, by random choice,
two incompatible \m s at the same
time, is actually an EPR-type
disentanglement experiment,
closely related to the micromaser
experiment discussed in the first part.\\

\noindent {\bf Keywords} Real
experiment. Delayed-choice erasure.
Distant measurement. Random-choice
EPR-type disentanglement.
Detector as \QMl system\\

{\footnotesize \rm \noindent
\rule[0mm]{4.62cm}{0.5mm}

\noindent F. Herbut (mail)\\
Serbian Academy of Sciences and
Arts, Knez Mihajlova 35, 11000
Belgrade, Serbia\\
e-mail: fedorh@sanu.ac.rs}\\

\pagebreak

{\bf \noindent 1 Introduction}\\

\noindent Scully and Dr\"{u}hl
published a thought experiment on
erasure in 1982 [1], which, 18
years later Kim, Yu, Kulik, Shih,
and Scully reported to have
performed in an inessentially
changed way \cite{Kim}. In this
article we investigate the
experiment because it realizes
several provoking and baffling
fundamental quantum-mechanical
ideas: (i) delayed choice (in the
sense of Wheeler \cite{Wheeler}),
(ii) erasure, (iii) erasure in
part of the state, (iv)
delayed-choice erasure in
\noindent the sense of Scully (or
after-detection erasure), (v)
random choice of particle-like or
wave-like behavior after
detection, and finally, (vi)
EPR-type disentanglement.

The authors lean on Glauber's
second-quantization theory for
precise quantitative \QMl
predictions, which turn out well
confirmed by the experiment. (For
references to Glauber's theory and
references to earlier work see the
article of Kim et al. \cite{Kim}.)

One of the first attempts to perform a
real erasure experiment \cite{Kwiat}
also presented its theoretical part in
second quantization. However, it has
turned out that first-quantization \QMl
insight \cite{FHMV} is feasible and
useful.

A \QMl analysis of the Kim et al.
article [2] is presented in this
paper in order to provide insight,
shed more light, and help to
demystify the mentioned puzzling
\QMl ideas. At last but not at
least, this study, along with the
preceding ones along parallel
lines (Ref-s [5] and
\cite{FHScully1}), should
hopefully help to understand in
what direction one should look for
an objective, preparation- and
observation-independent \qm .\\

\vspace{0.5cm}

{\bf \noindent 2 Basic Idea of the Kim
et al. Experiment and Questions}\\

\noindent A quantum eraser
experiment very close to and
somewhat simpler in details than
the experiment of Kim et al.
\cite{Kim} itself is illustrated
in the Fig. (The notation is in
accordance with that in the first
part of this study
\cite{FHScully1}.)

\begin{figure}

\includegraphics[width=10cm]{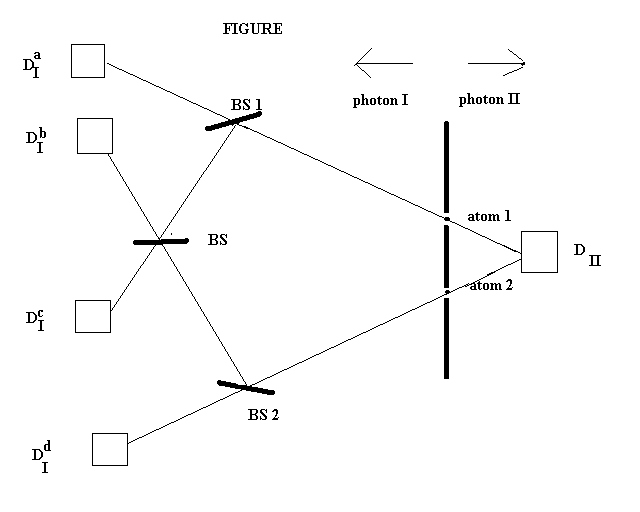}
\caption{The two-photon 'two-slit'
random-choice and delayed-choice
erasure experiment outlined in
section 2.}
\end{figure}

Two atoms labeled by \$1\$ and
\$2\$ (counterparts of the two
slits in Young's experiment
\cite{Young}) are excited by a
weak laser pulse. A pair of
entangled quanta, photon \$I\$ and
photon \$II\$, is then emitted
from either atom \$1\$ or atom
\$2\$ (coherently added
possibilities) by atomic cascade
decay (emitting photons \$I\$ and
\$II\$). Photon \$II\$,
propagating to the right, is
registered by detector \$D_{II}\$,
which can be scanned by a step
motor along its \$x\$ axis for the
observation of interference
fringes.

Photon \$I\$ propagates to the
left. If the pair is generated in
atom \$1\$, photon \$I\$ will
follow path \$1\$ (see the Fig.)
meeting beam splitter \$BS1\$ with
50\% chance of being reflected or
transmitted. If the pair is
generated in atom \$2\$, photon
\$I\$ will follow path \$2\$
meeting the beam splitter \$BS2\$
with 50\% chance of being
reflected or transmitted. In case
of the 50\% chance of being
transmitted at either \$BS1\$ or
\$BS2\$, photon \$I\$ is detected
by either detector \$D_I^a\$ or
\$D_I^d\$. The registration of
\$D_I^a\$ or \$D_I^d\$ provides
the {\it which-path information}
(path \$1\$ or path \$2\$
respectively) of photon \$I\$, and
this in turn provides the
which-path information for photon
\$II\$ due to the entanglement
nature of the two-photon state
generated by the atomic cascade
decay.

Given a reflection at either
\$BS1\$ or \$BS2\$, photon \$I\$
continues its \$1\$ or \$2\$ path
respectively to meet another
(central) \$50\%-50\%\$ beam
splitter \$BS\$, and then,
possibilities \$1\$ and \$2\$
having {\it interfered}, photon
\$I\$ is detected by either
detectors \$D_I^b\$ or \$D_I^c\$
shown in the Fig. The triggering
of detectors \$D_I^b\$ or
\$D_I^c\$ erases the which-path
information of photon \$II\$, and
creates the {\it which-coherence}
information.

The random choice takes place in the
beam splitters \$BS1(2)\$. As to the
delayed-choice (in the sense of
Scully), the which-path or both-path
(random) choice in the beam splitters
\$BS1\$ and \$BS2\$ is delayed compared
to the detection of photon \$II$ in
detector \$D_{II}\$.\\

There are some {\it questions} that
come to mind.

(i) If the quantum correlations in a
bipartite state, like that of the
two-photon system at issue, are such
that one particle contains which-path
information on the other, then there is
no coherence in the single photon state
(hence, one cannot detect interference)
as it is well known. Can this be made
obvious in the Kim et al. experiment?

(ii) It is known that pure-state
entanglement is due to coherence in the
state of the composite system, i. e.,
it stems from superposition of
orthogonal uncorrelated bipartite
states. How is the which-path
information erased, and, particularly,
how does the {\it coherence descend}
from the bipartite system to the
subsystem of photon-$II\$?

(iii) The behavior of the (improper
\cite{D'Espagnat}) ensemble of photons
\$II\$ is described by its state, i.
e., by the reduced density operator
\$\rho_{II}\$, of photon \$II\$. It is
known that this state cannot be
influenced by whatever happens to the
photon-$I\$ partner (improper) ensemble
alone (as long as there is no
interaction between the two photons).
Can one see this in the description of
the experiment?

(iv) Can a gradual increase in
complexity of the concepts involved be
displayed? There is the delayed choice
in the sense of Wheeler \cite{Wheeler},
random choice between which-path and
coherence information, Wheeler's
delayed choice with erasure, and,
finally, delayed choice in the sense of
Scully with erasure in part of the
state.

(v) It is known that quantum
correlations can not provide
signalling. This means that whatever
happens to the individual photon \$I\$,
no change is detectable on its partner,
photon \$II\$. How is erasure feasible
in view of this fact?

(vi) Photon \$I\$ 'makes the delayed
choice' by {\it randomly} being either
transmitted in \$BS1(2)\$ or being
reflected there. The photon-$II\$
partner has {\it before that been
detected} in detector \$D_{II}\$. How
can the one-path or interference origin
of this localization be decided later,
after the photon has been detected?\\

Next we are going to give a precise
first-quantization form to the above
outline of a physical picture in order
to enable one to answer the questions
and gain generally more insight in
the experiment.\\

\vspace{0.5cm}

{\bf \noindent 3. Quantum-mechanical
Description}\\

\noindent In order to answer the
questions posed, we transform now
the verbal description of the
experiment from the preceding
section into a \QMl two-photon
state vector following the example
of the simple \QMl description of
the Mach-Zehnder interferometer
\cite{Bub} (pp. 189-190).

One should have in mind that the
"path" state of the photon is
multiplied by the imaginary unit
when reflected on a beam splitter,
and is unchanged when transmitted
(\cite{Bub}, p. 189). Further, we
denote by \$\ket{0}_{D_{II}}\$ the
state vector of the localization
detector \$D_{II}\$ of the second
photon at the beginning of the
experiment, and by \$U_{II
D_{II}}\$ the unitary evolution
operator containing the
interaction between photon \$II\$
and the localization detector in
the process of localization and
afterwards. Then, the two-photon
state vector in the experiment,
{\it after passage of photon \$I\$
through the central beam splitter}
\$BS\$ (see the Fig.), but before
it reaches any of the detectors
\$D_I^q,\enskip q=a,b,c,d\$,
having in mind the above outline
of the experiment, reads as
follows. (We write it down, then
explain in detail.)

$$\Big|(1,2)\rightarrow
D_I^a,D_I^b,D_I^c,D_I^d\Big>_{I,II}=
(1/2)\Big|1\rightarrow D_I^a\Big>_I
\bigg\{U_{IID_{II}}\Big(\Big|1\Big>_{II}
\Big|0\Big>_{D_{II}}\Big)\bigg\}\quad
+$$
$$(1/2)\bigg\{(1/2)^{1/2}i
\bigg[i\Big|1\rightarrow D_I^b\Big>_I+
\Big|1\rightarrow
D_I^c\Big>_I\bigg]\bigg\}
\bigg\{U_{IID_{II}}\Big(
\Big|1\Big>_{II}\Big|0\Big>_{D_{II}}
\Big)\bigg\}\quad +$$
$$(1/2)\bigg\{
(1/2)^{1/2}i\bigg[\Big|2\rightarrow
D_I^b\Big>_I+i\Big|2\rightarrow
D_I^c\Big>_I\bigg]\bigg\}\bigg\{
U_{IID_{II}}\Big(\Big|2\Big>_{II}
\Big|0\Big>_{D_{II}}\Big)\bigg\}\quad
+$$
$$(1/2)\Big|2\rightarrow
D_I^d\Big>_I\bigg\{U_{IID_{II}}
\Big(\Big|2\Big>_{II}\Big|0\Big>_{D_{II}}
\Big)\bigg\}.\eqno{(1)}$$

There are \$4\$ (coherently added)
possibilities for photon \$I\$
expressed by the rhs of (1). It
may be emitted from atom \$1\$.
Then it may be transmitted through
or reflected from beam splitter
BS1. These are the first two terms
in (1). The last two terms cover
the symmetric case: the
possibility that photon \$I\$ is
emitted from atom \$2\$.

In the moment of our description
photon \$II\$ has already been
absorbed in detector \$D_{II}\$,
and the absorbed photon with the
detector evolves (in some
interacting way). Though, this
seems to occur locally, i. e.,
independently of what happens to
photon \$I\$, we must distinguish
the two basic possibilities:
photon \$II\$ emitted by atom
\$1\$ and photon \$II\$ emitted by
atom \$2\$, because we make this
distinction for photon \$I\$, and
photons \$I\$ and \$II\$ are
emitted together (we disregard the
small delay due to the cascade
emission from the same atom).
Hence, the first two terms in (1)
have one and the same tensor
factor
\$\bigg\{U_{IID_{II}}\Big(\Big|1\Big>_{II}
\Big|0\Big>_{D_{II}}\Big)\bigg\}\$,
and the last two terms have the
factor of the same form in which
\$1\$ is replaced by \$2\$.

The four possibilities exclude
each other. Hence the terms are
orthogonal. They are equally
probable; each has probability
\$1/4\$. The first term is the
simplest because, having been
transmitted through the beam
splitter BS1, the spatial state
vector \$\Big|1\rightarrow
D_I^a\Big>_I\$ suffers no change
of its phase factor.

The second term describes photon
\$I\$ after it has passed the
central beam splitter \$BS\$. The
photon can be reflected from it.
This gives \$i\Big|1\rightarrow
D_I^b\Big>_I\$. The other, equally
probable, possibility is passing
through \$BS\$, which results in
\$\Big|1\rightarrow
D_I^c\Big>_I\$. The two
possibilities exclude each other
(the corresponding terms are
orthogonal) and they give the
first-photon state vector in the
large brackets in the second term
of (1). It is multiplied by \$i\$
because the two possibilities in
it take place after reflection
from \$BS1\$.

As mentioned, the third and fourth
terms describe the symmetric cases
stemming from the (coherently
added) possibility that photon
\$I\$ was emitted from the second
atom.

Actually, there should also be the
(coherently added) possibility
that photon \$I\$ misses the beam
splitters \$BS1(2)\$ etc. But this
component of the bipartite state
vector is left out (projected out)
because it is irrelevant in the
experiment.\\

To answer the questions from the
preceding section, we need,
besides the state vector (1), also
the state (reduced density
operator) of photon \$II\$. This
is easy to evaluate if one
rewrites (1) in the form of an
expansion in an ortho-normal basis
for photon \$I\$. But first, we
can simplify (1).

One actually has
\$\Big|1\rightarrow
D_I^r\Big>_I=\Big|2\rightarrow
D_I^r\Big>_I,\enskip r=b,c\$
because, after passage through the
central beam splitter, photon
\$I\$ cannot 'remember' where it
has come from. Therefore, we can
write instead of
\$\ket{q\rightarrow
D_I^r}_I\enskip q=1,2\enskip
r=b,c\$ \$\ket{\rightarrow
D_I^r}_I\$. One can also replace
\$\ket{1\rightarrow D_I^a}_I\$ and
\$\ket{2\rightarrow D_I^d}_I\$ by
\$\ket{\rightarrow D_I^a}_I\$ and
\$\ket{\rightarrow D_I^d}\$
respectively because detector
\$D_I^a\$ can be reached only if
photon \$I\$ is emitted from atom
\$1\$ and symmetrically (we omit
the redundant information).

After omission of \$1(2)\$ in the
state vector of photon \$I\$ in
(1) as explained, relation (1) can
be rearranged so that we have an
expansion in the ortho-normal
basis \$\{\Big|\rightarrow
D_I^r\Big>_I:r=a,b,c,d\}\$ of
photon \$I\$ as easily seen.

$$\Big|(1,2)\rightarrow
D_I^a,D_I^b,D_I^c,D_I^d\Big>_{I,II}=$$
$$(1/2)\Big|\rightarrow
D_I^a\Big>_I\bigg\{U_{IID_{II}}\Big(
\Big|1\Big>_{II}\Big|0\Big>_{D_{II}}
\Big)\bigg\}\quad +$$
$$(1/2)\Big|\rightarrow D_I^b\Big>_I
U_{IID_{II}}\bigg\{\bigg[(1/2)^{1/2}
\bigg(-
\Big|1\Big>_{II}+i\Big|2\Big>_{II}
\bigg)\bigg]\Big|0\Big>_{D_{II}}
\bigg\}\quad +$$
$$(1/2)\Big|\rightarrow D_I^c\Big>_I
U_{IID_{II}}\bigg\{\bigg[
(1/2)^{1/2}\bigg(i
\Big|1\Big>_{II}-\Big|2\Big>_{II}
\bigg)\bigg]\Big|0\Big>_{D_{II}}
\bigg\}\quad +$$
$$(1/2)\Big|\rightarrow
D_I^d\Big>_I\bigg\{U_{IID_{II}}\Big(
\Big|2\Big>_{II}\Big|0\Big>_{D_{II}}
\Big)\bigg\}.\eqno{(2)}$$

The state (reduced density operator)
$$\rho_{II}\equiv\tr_I\bigg( \Big|(1,2)
\rightarrow
D_I^a,D_I^b,D_I^c,D_I^d\Big>_{I,II}
\Big<(1,2)\rightarrow
D_I^a,D_I^b,D_I^c,D_I^d\Big|_{I,II}
\bigg)$$ of photon \$II\$ can now
easily be evaluated from (2)
(because all 'off-diagonal' terms
give zero due to the orthogonality
of the first-photon basis).

$$\rho_{II}=(1/4)\bigg\{U_{IID_{II}}
\Big(\Big|1\Big>_{II}\Big|0\Big>_{D_{II}}
\Big)\bigg\}\bigg\{\Big(
\Big<1\Big|_{II}\Big<0\Big|_{D_{II}}
\Big)U_{IID_{II}}^{\dag}\bigg\}\quad
+$$
$$(1/4)U_{IID_{II}}\bigg\{\bigg[
(1/2)^{1/2}\Big(-\Big|1\Big>_{II}+i
\Big|2\Big>_{II}\Big)\bigg]
\Big|0\Big>_{D_{II}}\bigg\}\times$$
$$\bigg[(1/2)^{1/2}\Big(-\Big<1\Big|_{II}
-i\Big<2\Big|_{II}\Big)\bigg]
\Big<0\Big|_{D_{II}}\bigg\}
U_{IID_{II}}^{\dag}\quad +$$
$$(1/4)U_{IID_{II}}\bigg\{\bigg[
(1/2)^{1/2} \Big(i\Big|1\Big>_{II}-
\Big|2\Big>_{II}\Big)\bigg]
\Big|0\Big>_{D_{II}}\bigg\}\times
$$
$$\bigg\{\bigg[(1/2)^{1/2}
\Big(-i\Big<1\Big|_{II}
-\Big<2\Big|_{II}\Big)\bigg]
\Big<0\Big|_{D_{II}}\Bigg\}
U_{IID_{II}}^{\dag}\quad +$$
$$(1/4)\bigg\{U_{IID_{II}}\Big(
\Big|2\Big>_{II}\Big|0\Big>_{D_{II}}
\Big)\bigg\}\bigg\{\Big(
\Big<2\Big|_{II}\Big<0\Big|_{D_{II}}
\Big)U_{IID_{II}}^{\dag}\bigg\}.
\eqno{(3)}$$

Multiplying out the terms, one obtains

$$\rho_{II}=(1/2)\bigg\{U_{IID_{II}}
\Big(\Big|1\Big>_{II}\Big|0\Big>_{D_{II}}
\Big)\bigg\}\bigg\{
\Big<1\Big|_{II}\Big<0\Big|_{D_{II}}
\Big)\bigg\}U_{IID_{II}}^{\dag}\quad
+$$ $$(1/2)\bigg\{U_{IID_{II}}\Big(
\Big|2\Big>_{II}\Big|0\Big>_{D_{II}}
\Big)\bigg\}\bigg\{\Big(
\Big<2\Big|_{II}\Big<0\Big|_{D_{II}}
\Big)U_{IID_{II}}^{\dag}\bigg\}.
\eqno{(4)}$$

The time variable in the unitary
evolution operator \$U_{IID_{II}}\$ has
been suppressed. As it has been stated,
its value is some instant after photon
\$I\$ has passed the central beam
splitter \$BS\$, but has not yet
reached any of the four detectors
\$D_I^a$-$D_I^d\$ (or, possibly, has
already been detected in the first or
in the last of them).

If the experiment were a {\it
simple erasure} (before-detection)
one, then one would still have
lack of interaction between photon
\$II\$ and the localization
detector \$D_{II}\$ at the moment
in question:
\$U_{IID_{II}}=U_{II}U_{D_{II}}\$.
Hence, detector \$D_{II}\$ could
be omitted (moved from the
'object' of description to the
'subject'). If, on the contrary,
we have delayed-choice (or
after-detection) erasure, then we
have interaction, and the
description as it stands is
essential (cf insight in
delayed-choice erasure in part I
\cite{FHScully1}).\\

One should note that the four
terms in (1) are coherently mixed
(the terms are superposed). It is
of crucial importance to preserve
this coherence during the
experiment. Nowadays much thinking
goes on about the problem how to
counteract decoherence (see e. g.
\cite{Agarwal}), which destroys
coherence.\\

\vspace{0.5cm}

{\bf \noindent 4 Answers}\\

\noindent Relations (1) and (4)
give answer to {\it question} (i).
The former makes it obvious that
orthogonal states of photon \$I\$
in the first two terms on the one
hand and the last two terms on the
other 'mark' or distinguish the
'being emitted from atom \$1$' and
'being emitted from atom \$2$'
respective states of photon \$II\$
(latent 'which-path' information).
Relation (4) then makes it evident
that this has the consequence of
'suppressing' the coherence in the
state of photon \$II$.

Note that the much-used term {\it
'erasure'} is not meant to be a
synonym for this 'suppression'. On
the contrary, it denotes {\it
elimination of the described
mechanism of 'suppression'}, which
is still present in (1) or (2).
The very detection of the beam of
photons \$I\$ in the detectors
\$D_I^r\enskip r=a,b,c,d\$, turns
(2) into a (proper) mixture (of
the 4 component states in (2)).
Then, also \$\rho_{II}\$ given by
(3) is a proper mixture containing
two
coherence states (terms \$2\$ and \$3\$).\\

Answer to {\it question} (ii)
follows from the fact that the
central beam splitter \$BS\$ (cf
the Fig.) treats the components of
photon \$I\$ coming from atoms
\$1\$ and \$2\$ {\it differently
only by a phase difference of
\$(\pi /2)\$} (multiplication by
\$i\$). This makes possible {\it
erasure}, so that, upon detection
in \$D_I^r,\enskip r=a,b,c,d\$,
half of the ensemble of photons
\$II\$ displays interference (cf
relation (3)).

Note that state decomposition (3), as
it stands, is only a mathematical
relation. The second and third
subensembles of the ensemble of photons
\$II\$ in it are {\it not defined
physically on the local, photon-$II$,
level}. They are defined only distantly
(or, one might say, globally) on part
of photon \$I\$ propagating towards the
detectors
\$D_I^r,\enskip r=b,c$.\\

{\it Question} (iii) is answered
in the affirmative by comparing
relations (2) and (4): in spite of
the erasure in part of the
two-photon state vector displayed
in (2), locally, i. e., in the
entire ensemble of photons \$II\$,
there is no change induced (as
obvious
in (4)).\\

To answer {\it question} (iv), we can
begin by drastically downgrading the
Kim et al. experiment, and then by
upgrading it in steps.

{\bf a)}. Let us imagine that one,
so to say, 'by hand' either just
removes the beam splitters
\$BS1(2)\$ or, as an alternative,
replaces them by mirrors. Further,
we imagine that the choice between
these two possibilities is made
{\it after} the photons begin to
propagate from the atoms (the
'slits'). Thus, the choice of a
particle-like or a wave-like
experiment would be {\it delayed}
with respect to the moment of
preparation (or beginning of the
experiment).

This would constitute a genuine Wheeler
delayed-choice experiment (though
upgraded from the original one-photon
case \cite{Wheeler} to a two-photon
one).

{\bf b)}. In the Kim et al. experiment
the 'by hand' choice is replaced by a
{\it random} mechanism (transmission or
reflection on the beam splitters
\$BS1(2)\$). This upgrading of
Wheeler's idea is extremely important
on two counts:

{\bf (i)} Let us remember the
famous wave-particle duality form
of Bohr's complementarity
principle and the rebellions
against its claim of universal
validity (see the impressive work
of Ghose and Home in \cite{Home},
and the references therein). The
random choice mingles
particle-like and wave-like
behavior in one experiment, thus
giving support to the mentioned
rebellions.

{\bf (ii)} At the very beginning of the
first erasure paper of Scully et al.
[1] the authors say:

\begin{quote}
"... {\it the role of the observer}
lies at the heart of the problem of
measurement and state reduction in \qm
." (Italics by F. H.)
\end{quote}

A few lines lower they say:

\begin{quote}
"... Wheeler has pointed out that {\it
the experimentalist} may delay his {\it
decision} as to display wave-like or
particle-like behavior in a light beam
long after the beam has been split by
the appropriate optics." (Italics by F.
H.)
\end{quote}

The random-choice property of the Kim
et al. experiment (beam splitters
\$BS1(2)\$) has replaced the 'decision'
of the 'observer' or the 'experimenter'
by an automatic step in the experiment.
This goes a long way in warning against
the fallacy of overestimating the role
of the human observer.

The Kim et al. experiment performs
a {\it partial-state erasure}
having the two-photon state in
mind. This is not partial erasure.
The erasure is complete, but it
takes place only in a part of the
two-photon state. Since states are
ensembles in the laboratory, one
experimentally deals with two
sub-ensembles: the one 'seen' in
the coincidence of the detectors
\$D_I^r\$ and \$D_{II},\$
\$r=a,d,\$ ('which path') and the
other 'seen' in the
\$D_I^r,D_{II}, \enskip r=b,c\$
coincidence ('which
interference').

{\bf c)} Kim et al. actually performed
a delayed-choice (in the sense of
Scully) or after-detection (of photon
\$II\$) experiment in which this
detection was performed so early in the
experiment, that photon \$I\$ has not
yet reached the beam splitters
\$BS1(2)\$, i. e., the random choice
has not yet taken place. (This seems to
put a dramatic
emphasis on question {vi).)\\

I believe that the answer to {\it
question} (v) has now become clear. The
fact that no signal can be sent from
photon \$I\$ to photon \$II\$ is based
on the circumstance that the {\it
entire state} (ensemble) \$\rho_{II}\$
of photons \$II\$ (cf (4)) cannot be
distantly manipulated on account of
entanglement. Whatever change photon
\$I\$ undergoes due to some local
interaction, this does not influence
the state \$\rho_{II}\$ of photon
\$II.\$ But this is not so if one makes
coincidence measurements on photon
\$I\$ and photon \$II.\$ Then distant
(or global) influence does show up and
erasure can appear.\\

Answer to {\it question} (vi) was
given in part I \cite{FHScully1}:
the combined photon-$II$-$D_{II}\$
system is analogous to the
photon-$II$ system alone as far as
the definite-way or coherence
states are concerned. Therefore,
there is no question of 'acting
backwards in time' or any other
mystification.\\

\vspace{0.5cm}

{\bf \noindent 5 The Experiment is an
EPR-type Disentanglement}\\

\noindent Now we proceed to a deeper
layer of physical insight in the
random-choice and delayed-choice
erasure experiment that we investigate:
we view it as an EPR-type
disentanglement in which both mutually
incompatible distant \m s are performed
simultaneously.

The random EPR-type
disentanglement interpretation of
the experiment can, actually, be
seen from (2) and (3). Namely, it
is evident that the first and the
last terms in both equations refer
to distant 'which-path' relation,
and the second and third terms in
them describe 'which-interference'
relations, both in a latent way
until the corresponding detector
is reached. It is clear from (3)
and (2) that the improper ensemble
of photons \$II\$ is seen as
broken up into the mentioned two
(improper) sub-ensembles
distantly, i. e., due to a
relevant difference in the
photon-$I-$ partners. Locally,
each individual photon \$II\$ is
in the same state \$\rho_{II}\$
given by (4).

This decomposition of the ensemble
is of the type
\$\rho_{II}=(1/2)\rho_{II}+(1/2)
\rho_{II}\$. It is one of the
beauties of the experiment.
Namely, both the 'which-path' part
and the 'which-interference' part,
taken separately, look like the
two separate experiments which
make up the EPR-type
disentanglement, and which are in
most other experiments performed
alternatively at the will of the
experimenter. Here both
experiments are performed together
on account of the random-choice
function of the first beam
splitters \$BSq,\enskip q=1,2\$.

It is also worth mentioning that
one has two interferences
$$(1/2)^{1/2}\bigg[-U_{IID_{II}}\bigg(
\Big|1\Big>_{II}\Big|0\Big>_{D_{II}}\bigg)
+iU_{IID_{II}}\bigg(\Big|2\Big>_{II}
\Big|0\Big>_{D_{II}}\bigg)\bigg]$$
and
$$(1/2)^{1/2}\bigg[iU_{IID_{II}}\bigg(
\Big|1\Big>_{II}\Big|0\Big>_{D_{II}}\bigg)
-U_{IID_{II}}\bigg(\Big|2\Big>_{II}
\Big|0\Big>_{D_{II}}\bigg)\bigg]$$
(cf (2)), which are 'opposite' in
the sense that the corresponding
(pure-state) density matrices add
up into \$\rho_{II}\$ given by (4)
(cf the second and third terms in
(3),
which add up into (4)).\\

All that remains to be done is to
prove a formal claim made in the
first part \cite{FHScully1} of
this study. It was shown there
that there are two {\it simple
coherence bases} (in the Schmidt
canonical expansion relevant for
EPR-type disentanglement)
\$\ket{\pm }_I\equiv (1/2)^{1/2}
\Big(\ket{1}_I\pm\ket{2}_I \Big)\$
(equation (10) there) and
\$\ket{\pm i}_I\equiv (1/2)^{1/2}
\Big(\ket{1}_I\pm
i\ket{2}_I\Big)\$ (equation (11)
there). It was, further, shown
that the first simple coherence
bases are realized in the
micro-maser experiment
\cite{Scully1}, and it was claimed
that the second simple coherence
basis finds realization in the
experiment of Kim et al. that is
the object of investigation in
this article.

If we project out and renormalize
the 'which-coherence' part in (2),
and move some numerical factors
from the second tensor factor to
the first, we obtain
$$\ket{(1,2)\rightarrow
D_I^b,D_I^c}_{I,II}=$$
$$(1/2)^{1/2}
\Bigg(-\Big|\rightarrow
D_I^b\Big>_I
U_{IID_{II}}\bigg\{\bigg[(1/2)^{1/2}
\bigg(\Big|1\Big>_{II}-
i\Big|2\Big>_{II}
\bigg)\bigg]\Big|0\Big>_{D_{II}}
\bigg\}\quad +$$
$$i\Big|\rightarrow D_I^c\Big>_I
U_{IID_{II}}\bigg\{\bigg[
(1/2)^{1/2}\bigg(\Big|1\Big>_{II}+
i\Big|2\Big>_{II}
\bigg)\bigg]\Big|0\Big>_{D_{II}}
\bigg\}\quad\Bigg).\eqno{(5)}$$

Next we find out how to express
the orthonormal basis vectors
\$-\ket{\rightarrow D_i^b}_I,\$
\$i\ket{\rightarrow D_I^c}_I\$
appearing in (5) in terms of the
state vectors
\$\ket{1}_I,\enskip\ket{2}_I\$,
which (by definition) denote the
component states of photon \$I\$
propagating from the respective
first beam splitters \$BSq,\enskip
q=1,2,\$ towards the central beam
splitter \$BS\$ (cf the Figure).
The state vectors
\$\ket{1}_I,\enskip\ket{2}_I\$ are
the normalized relevant
projections of the initial
corresponding state vectors
expressing propagation from the
atoms to the first beam splitters.
Their definition corresponds to
that of the coherence component
(5) of the two-photon state
vector, which is analogously
projected out (and renormalized)
from (2).

Let the unitary operator \$U_I\$
stand for the evolution of the
first photon from these component
state vectors \$\ket{q}_I,\enskip
q=1,2,\$ to the respective states
that appear after passing the
central beam splitter \$BS\$
(where \$\ket{\rightarrow D_I^r},
\enskip r=b,c\$ are defined).

We make use of the equality
\$\Big|1\rightarrow
D_I^r\Big>_I=\Big|2\rightarrow
D_I^r\Big>_I,\enskip r=1,2\$ (cf
beneath (1)) again. Then, one can
see from relation (1) that
$$U_I\ket{1}_I=(1/2)^{1/2}
\Big(-\ket{\rightarrow D_I^b}_I+
i\ket{\rightarrow
D_I^c}_I\Big),\eqno{(6a)}$$
$$U_I\ket{2}_I=(1/2)^{1/2}
\Big(i\ket{\rightarrow D_I^b}_I-
\ket{\rightarrow
D_I^c}_I\Big),\eqno{(6b)}$$

Solving (6a) and (6b) for
\$\ket{\rightarrow
D_I^r}_I,\enskip r=b,c\$, one
obtains
$$-\ket{\rightarrow D_I^b}_I=
(1/2)^{1/2}\Big[(U_I
\ket{1}_I)+i(U_I\ket{2}_I)\Big],
\eqno{(7a)}$$ and
$$i\ket{\rightarrow D_I^c}_I=
(1/2)^{1/2}\Big[(U_I
\ket{1}_I)-i(U_I\ket{2}_I\Big)\Big].
\eqno{(7b)}$$

If we replace in (5) the basis
vectors \$-\ket{\rightarrow
D_i^b}_I,\$ \$i\ket{\rightarrow
D_I^c}_I\$ with the expressions
given by (7a) and (7b)
respectively, we obtain
$$\ket{(1,2)\rightarrow
D_I^b,D_I^c}_{I,II}=
\Big\{(1/2)^{1/2}\Big[(U_I
\ket{1}_I)+i(U_I\ket{2}_I)
\Big]
\Big\}\otimes$$
$$\bigg\{(1/2)^{1/2}
\bigg[\Big(U_{IID_{II}}(\Big|1\Big>_{II} \Big|0\Big>_{D_{II}})\Big)-
i\Big(U_{IID_{II}}(\Big|2\Big>_{II} \Big|0\Big>_{D_{II}})\Big)\bigg]
\bigg\}\quad +$$
$$\Big\{(1/2)^{1/2}\Big[(U_I
\ket{1}_I)-i(U_I\ket{2}_I)\Big]
\Big\}\otimes$$
$$\bigg\{(1/2)^{1/2}
\bigg[\Big(U_{IID_{II}}(\Big|1\Big>_{II})
\Big|0\Big>_{D_{II}})\Big)+
i\Big(U_{IID_{II}}(\Big|2\Big>_{II}
\Big|0\Big>_{D_{II}})\Big)\bigg]
\bigg\}.\eqno{(8)}$$

Equation (8) is a Schmidt
canonical expansion, essentially,
in the mentioned second simple
basis \$\ket{\pm i}_I\equiv
(1/2)^{1/2}\Big(\ket{1}_I\pm
i\ket{2}_I\Big)\$.

To understand better the state
vectors appearing in (8), we write
down also the 'which-path' part of
(2). For comparison with (8), let
us introduce the state vectors
\$\ket{q}'_I,\enskip q=1,2,\$ as
the components that are
transmitted through the beam
splitters \$BSq\$, and for their
further evolution we utilize
\$U'_I\$, under the action of
which they become
\$\ket{\rightarrow
D_I^r}_I,\enskip r=a,d\$
respectively. We obtain
$$\ket{(1,2)\rightarrow
D_I^a,D_I^d}_{I,II}=
(U'_I\ket{1}'_I)\Big(U_{IID_{II}}
(\ket{1}_{II}\ket{0}_{D_{II}})\Big)\quad
+$$
$$(U'_I\ket{2}'_I)\Big(U_{IID_{II}}
(\ket{2}_{II}\ket{0}_{D_{II}})
\Big).\eqno{(9)}$$

If the lhs of (8) would equal that
of (9), and if one could drop the
prims in (9), i. e., if one could
write
\$U'_I\ket{q}'_I=U_I\ket{q}_I,\enskip
q=1,2,\$ then (8) and (9) would be
a true parallel to the
'which-path' and
'which-interference' EPR-type
disentanglement in the micromaser
discussed in part I
\cite{FHScully1}, only that
instead of the first, we would
have the second simplest basis (as
explained above).

In the real experiment that we are
discussing one must pay a price
for having the two complementary
disentanglements in one
experiment. Namely, one must by
projection and renormalization
decompose (2) into (9) and (8)
because they are two distinct
parts of the same experiment (not
two versions). Correspondingly,
one has
\$\ket{q}'_I\not=\ket{q}_I,\enskip
q=1,2\$ on account of the two
distinct projections (and
renormalizations) of the component
states.

Nevertheless, having these peculiarities in mind,
the claim from
part I has been shown to be, essentially,
valid.\\

\pagebreak

\noindent
{\bf 6 Concluding Remarks}\\

\noindent  The experiment
discussed in this article provides
us with a clear understanding of
the distinction of 'potential' and
'actual' in the usual sense of
these words (cf remark F in the
first part of this study
\cite{FHScully1}).

Let us think of the experiment at
issue as if it were performed in
the photon-by-photon version. Then
all the coherent possibilities
(terms in (1) or (2)) are the
photon's realities though still in
a relative sense with respect to
the preparator. We call them
'potential' with respect to our
subjective choices of highlighting
parts of it. 'Actuality' comes to
the fore when we consider a
detection coincidence, e. g.,
\$D_I^b\$ and \$D_{II}^n\$ (see
part I [6] for \$D_{II}^n\$).
Then, taking photon by photon, and
considering different values of
\$n\$, the fringes of the
corresponding interference pattern
come about. But this is no more
than a subjective highlighting of
part of the reality of the
experiment. The single photon in
reality has both the which-path
property and the coherence
property. (And by this it has both
which-path and both coherence
possibilities.)

This is analogous to the original
EPR situation \cite{EPR} with
position and linear momentum
disentanglement. But the present
experiment has a big mentioned
advantage. The original EPR
(thought) experiment could have
this advantage only if it
contained \M of position and
momentum in the same experiment.\\

The Kim et al. experiment is, in
the opinion of the present author,
perhaps the most accomplished
realization of an EPR-type
disentanglement. I believe that
this experiment is so important
from the foundational point of
view that it should be performed
both in the photon-by-photon
version, like e. g. the
Mach-Zehnder interferometer
experiments \cite{Aspect1},
\cite{Aspect2}, and in terms of
positive-rest-mass particles.\\

Jaynes writes \cite{Jaynes} (the last
passage in the web version):

\begin{quote}
"... it is pretty clear why present
quantum theory not only does not use -
it does not even dare to mention - the
notion of a "real physical situation".
Defenders of the theory say that this
notion is philosophically naive, a
throwback to outmoded ways of thinking,
and that recognition of this
constitutes deep new wisdom about the
nature of human knowledge. I say that
it constitutes a violent irrationality,
that somewhere in this theory the
distinction between reality and our
knowledge of reality has become lost,
and the result has more the character
of medieval necromancy than of science.
It has been my hope that quantum
optics, with its vast new technological
capability, might be able to provide
the experimental clue that will show
how to resolve these contradictions."
\end{quote}

Isn't it possible that Jaynes'
hope has, at least to some extent,
come true precisely on account of
the work of Scully et al.? I find
it hard to think of the Kim et al.
experiment \cite{Kim} in any other
terms than as a "real physical
situation". Its comprehension
suggests the RRUES interpretation
(
see part I). Then why not think
of reality, at least as far as
experiments are concerned, in this
way? Understanding experiments is
the natural springboard for
understanding nature.\\

\vspace{0.5cm}

{\bf \noindent ACKNOWLEDGEMENTS}
The author is indebted to the
unknown referees who, besides
useful suggestions, also helped
substantially to achieve a
presentation that is more
readable.


\vspace{0.5cm}

\begin{thebibliography}{99}

\bibitem{Scully,Druehl}
Scully, M. O., Dr\"{u}hl, K.: Quantum
eraser: A proposed photon correlation
experiment concerning observation and
"delayed-choice" in \qm. Phys. Rev. A
{\bf 25}, 2208-2213 (1982)

\bibitem{Kim}
Kim Y.-H., Yu R., Kulik S. P., Shih Y.,
Scully M. O.: Delayed "choice" quantum
eraser. Phys. Rev. Lett. {\bf 84}, 1-5
(2000)

\bibitem{Wheeler}
Wheeler J. A.: The 'past' and the
delayed-choice double-slit experiment.
In: Marlow A. R. ed. Mathematical
Foundations of Quantum Theory. Academic
Press, New York (1978), pp. 9-48

\bibitem{Kwiat}
Kwiat P. G., Steinberg A. M., Chiao R.
Y.: Observation of a "Quantum Eraser":
A Revival of Coherence in a Two-photon
Interference Experiment. Phys. Rev. A
{\bf 45}, 77297739 (1992)

\bibitem{FHMV}
Herbut F., Vuji\v{c}i\'{c} M.:
First-quantisation quantum-mechanical
insight into the Hong-Ou-Mandel
two-photon interferometer with
polarizers and its role as a quantum
eraser. Phys. Rev. A {\bf 56}, 1-5
(1997)

\bibitem{FHScully1}
Herbut F.: On EPR-type
Entanglement in the Experiments of
Scully et al.I The Micromaser Case
and Delayed-Choice Quantum
Erasure. Found. Phys. {\bf 38},
1046-1064 (2008).
ArXiv:quant-ph/0808.3176

\bibitem{Young}
T. Young, Trans. R. Soc. XCII {\bf
12}, 387 (1802).\\ A
quantum-mechanical discussion in
\cite{FHAJP92}

\bibitem{D'Espagnat}
D'Espagnat B.: Conceptual Foundations
of Quantum Mechanics. Second Edition.
W. A. Benjamin, Inc., Reading,
Massachusetts (1976). Subsection 7.2

\bibitem{Bub}
Bub J.: Interpreting the Quantum World.
Cambridge University Press, N. Y.
(1997)

\bibitem{Agarwal}
Agarwal, G.S., Scully, M.O.,
Walther, H: Inhibition of
Decoherence due to Decay in a
Continuum. Phys. Rev. Lett. {\bf
86}, 4271-4274 (2001)

\bibitem{Home}
Ghose P., Home D.: The two-prism
experiment and wave-aprticle duality of
light. Found. Phys. {\bf 26}, 943-953
(1996)

\bibitem{Scully1} Scully,
M.O., Englert B.-G., Walther, H.:
Quantum optical tests of
complementarity. Nature {\bf 351},
111-116 (1991)

\bibitem{EPR}
Einstein A., Podolsky B., Rosen
N.: Can quantum-mechanical
description of physical reality be
considered complete? Phys. Rev.
{\bf 47}, 777-780 (1935)

\bibitem{Aspect1}
Grangier P., Roger G., Aspect A.:
Experimental evidence for a photon
anricorrelation effect on a beam
splitter: A new light on
single-photon interferences.
Europhys. Lett. {\bf 1}, 173-179
(1986)

\bibitem{Aspect2}
Jacques V., Wu E., Grosshans F.,
Treussart F.,Grangier P., Aspect
A., Roch J.-F.: Delayed-choice
test of complementarity with
single photons. Archiv
quant-ph/0801.0979

\bibitem{Jaynes}
Jaynes, E. T.: in
Foundation of Radiation Theory and
Quantum Electronics, Barut A. ed.,
Plenum, 1980; also item 38 of the
Jaynes bibliography on the web


\bibitem{FHAJP92}
Herbut F.: Quantum interference viewed
in the framework of probability theory.
Am. J. Phys. {\bf 60}, 146-150 (1992)




\end{thebibliography}
\end{document}